\documentclass[a4paper,12pt]{article}  
\usepackage[dvips]{graphicx}
\usepackage{amssymb}
\usepackage{cite}
\newcommand{\goes}{\rightarrow} 
 
\newcommand{\GeV}{\; \mathrm{GeV}} 
\newcommand{\TeV}{\; \mathrm{TeV}} 
\newcommand{\lapproxeq}{\lower .7ex\hbox{$\;\stackrel{\textstyle  
<}{\sim}\;$}} 
\newcommand{\gapproxeq}{\lower .7ex\hbox{$\;\stackrel{\textstyle  
>}{\sim}\;$}} 
\newcommand{\stackdown}[2]{\lower 1.4ex\hbox{$\;\stackrel{\textstyle{#1}}  
{\scriptstyle{#2}}\;$}}
\newcommand{\beq}{\begin{equation}} 
\newcommand{\eeq}{\end{equation}} 
\newcommand{\bea}{\begin{eqnarray}} 
\newcommand{\eea}{\end{eqnarray}}
\newcommand{\lsp}{\tilde{\chi}}
\newcommand{\mlsp}{m_{\lsp}}

\newcommand{\relic}{\Omega_{\lsp}\,h_0^2} 
 
\newcommand{\etal}{\textit{et. al.}}
\newcommand{\almuon}{\alpha_{\mu}^{\mathrm{SUSY}}}       
\newcommand{\sxsec}{\sigma_{scalar}}                     
\newcommand{\bsga}{b \goes s \, \gamma}

\makeatletter 
\def\slash{\@ifnextchar[{\fmsl@sh}{\fmsl@sh[0mu]}} 
\def\fmsl@sh[#1]#2{%
  \mathchoice 
    {\@fmsl@sh\displaystyle{#1}{#2}}%
    {\@fmsl@sh\textstyle{#1}{#2}}%
    {\@fmsl@sh\scriptstyle{#1}{#2}}%
    {\@fmsl@sh\scriptscriptstyle{#1}{#2}}} 
\def\@fmsl@sh#1#2#3{\m@th\ooalign{$\hfil#1\mkern#2/\hfil$\crcr$#1#3$}} 
\makeatother 

\begin{document} 
\begin{titlepage} 
 
\begin{flushright} 
\parbox{4.6cm}{hep-ph/0211286\\
               MIFP-02-08\\
               ACT-02-10\\
               HEPHY-PUB 768/02} 
\end{flushright} 
\vspace*{5mm} 
\begin{center} 
{\large{\textbf {Updating the constraints to CMSSM from cosmology\\
and accelerator experiments 
}}}\\
\vspace{14mm} 
{\bf A.~B.\ Lahanas} $^{1}$ \,,
{\bf D.~V.~Nanopoulos} $^{{2}\; \dagger}$ \, and \, {\bf V.~C.~Spanos} $^{3}$  

\vspace*{6mm} 
 $^{1}$ {\it University of Athens, Physics Department,  
Nuclear and Particle Physics Section,\\  
GR--15771  Athens, Greece}

\vspace*{6mm} 
$^{2}$ {\it George P. and Cynthia W. Mitchell 
                Institute of Fundamental Physics, \\   
     Texas A \& M University, College Station,  
     TX~77843-4242, USA \\[1mm]
     Astroparticle Physics Group, Houston 
     Advanced Research Center (HARC),\\ 
     The Mitchell Campus, The Woodlands, TX 77381, USA  \\[1mm] 
     Chair of Theoretical Physics,  
     Academy of Athens,  
     Division of Natural Sciences, 28~Panepistimiou Avenue,  
     Athens 10679, Greece }

\vspace*{6mm}
$^{3}$ {\it  Institut f\"ur Hochenergiephysik der \"Osterreichischen Akademie
der Wissenschaften,\\
  A--1050 Vienna, Austria}
\end{center} 
\vspace*{15mm} 
\begin{abstract}
The recent data from E821 Brookhaven experiment
in conjunction with a new determination of the 
 hadronic vacuum polarization contribution
to the anomalous magnetic moment of muon,
put new bounds on the
parameters of the Constrained Minimal Supersymmetric Standard Model.
We study the impact this 
experimental information, along with 
the $b \goes s \, \gamma$ branching ratio and
light Higgs boson mass bound from LEP,  
to constrain regions of the model which are
consistent with the cosmological data.
The effect of these to Dark Matter
direct searches is also discussed.
  
\end{abstract} 
\vspace{15mm}
\noindent
\rule[0.cm]{11.6cm}{.009cm} \\     
\vspace{1mm} 
$^\dagger$ Invited talk given at the International Conference on Physics
Beyond The Standard Model: ``Beyond the Desert 02'', Oulu, Finland 2--7 June,
2002.
\end{titlepage} 

\newpage 
\baselineskip=18pt 

\section{Introduction}
There is plenty of observational cosmological evidence that 
indicates the strong need for Cold Dark Matter (DM).
Interestingly enough, one of the major and rather
unexpected predictions of Supersymmetry (SUSY), broken
at low energies $M_{SUSY} \thickapprox \mathcal{O}(1 \TeV) $,
while $R$-parity is conserved, is the existence of a stable, neutral
particle, the lightest neutralino  ($\lsp$), 
referred to as the LSP \cite{Hagelin}.
Such particle is an ideal candidate 
for the Cold DM in the Universe \cite{Hagelin}.
The latest data from the Cosmic
Microwave Background (CMB) radiation 
anisotropies \cite{cmb} not only favour
a flat ($k=0$ or $\Omega_0=1$),
inflationary Universe, but they  also determine a matter density
$\Omega_M h_0^2 \thickapprox 0.15 \pm 0.05$.
Subtracting from this the  baryon density
$\Omega_B h_0^2 \thickapprox 0.02$, and the rather tiny
neutrino density, one gets $ \Omega_{DM} h_0^2 = 0.13 \pm 0.05 $ for
the DM density.
The most recent detection of the polarization 
in the CMB strengthens 
further the case for  Cold DM \cite{dasi}. 
On the other hand SUSY is not only  indispensable in 
constructing consistent string
theories, but
it also seems unavoidable at low energies ($\sim 1 \TeV$)
if the gauge hierarchy problem is to
be resolved.
Such a resolution provides a measure of the SUSY
breaking scale $M_{SUSY} \thickapprox \mathcal{O}(1 \TeV) $. 
There is  indirect evidence for such a
low-energy supersymmetry breaking scale, from the unification
of the gauge couplings \cite{Kelley} and from the apparent lightness
of the Higgs boson as determined from precise electroweak measurements,
mainly at LEP \cite{EW}.

Recently the BNL E821 experiment \cite{E821} delivered
a new and more precise measurement for the  
anomalous magnetic moment of the muon
\bea
\alpha_\mu^{\mathrm exp}=11659203(8)\times 10^{-10} \, , 
\eea 
where $\alpha_\mu = (g_\mu-2)/2$.
One the other hand, a detailed calculation 
about  the hadronic vacuum polarization contribution
to this moment appeared \cite{thomas}\footnote{A similar 
calculation\cite{davier} based on low-energy $e^+e^-$ 
data drew similar conclusions.}.
This calculation, especially using inclusive data,
favours  smaller values of the hadronic vacuum polarization.
 As a result, the discrepancy between the the Standard Model (SM)
theoretical prediction  and the experimental value for  the  
anomalous magnetic moment of the muon, becomes significant \cite{thomas}
\bea
\delta \alpha_{\mu} = (361 \pm 106) \times 10^{-11} \, ,
\eea
which corresponds to a $3.3 \, \sigma$ deviation.
Combining this with the cosmological bound for supesymmetric
dark matter density, one restricts  considerably the parameter
space of  the  Constrained Minimal 
Supersymmetric Standard Model (CMSSM). 
In our analysis we have taken also into account some
other important constraints: the branching ratio for
the $\bsga$ transition and the light Higgs mass bound
$m_{h} \geq 113.5 \GeV$  provided by LEP \cite{LEP}.
Concerning the $\bsga$ branching ratio the $2\, \sigma$
bound $1.8 \times 10^{-4} < BR(\bsga) < 4.5 \times 10^{-4}$ is
used \cite{cleo}.

\section{Neutralino relic density}
In the large $\tan\beta$ regime 
the neutralino ($\lsp$) pair annihilation 
through $s$-channel
pseudo-scalar Higgs boson ($A$) 
exchange, leads to an enhanced annihilation cross sections
reducing significantly the relic 
density \cite{Drees}. 
The importance of this mechanism, in conjunction with the recent
cosmological data which favour small values of the DM
relic density,
has been stressed in \cite{LNS,LNSd}. The same mechanism has been
also invoked  \cite{Ellis} where it 
has been shown that it enlarges the cosmologically
allowed regions. 
In fact cosmology does not put severe upper
bounds on sparticle masses, and soft masses can be in the TeV region,
pushing up the sparticle mass spectrum to regions that might escape detection
 in future planned accelerators. 
Such an upper bound is imposed, however, by
the  ($g_\mu -2$) E821 data  and has been  
the subject of intense phenomenological 
study the last year \cite{ENO,LS,LNSd2,g-2,Leszek,Kneur}.
As was mentioned above, for large $\tan \beta$ 
the $\lsp \, \lsp \stackrel{A}{\goes} b \, \bar{b}$ or $\tau \, \bar{\tau}$ 
channel becomes the dominant annihilation mechanism. 
In fact by increasing $\tan \beta$ the mass $m_A$ decreases, while the
neutralino mass remains almost constant, if the other parameters are kept
fixed. Thus  $m_A$ is expected eventually to enter into the regime in which
it is close to the pole value $m_A\,=\, 2 m_{\lsp}$, and the
pseudo-scalar  Higgs exchange dominates.
It is interesting to point out that in a previous analysis of the direct
DM searches \cite{LNSd}, we had stressed 
that  the contribution of the $CP$-even Higgs bosons exchange
to the LSP-nucleon scattering cross sections increases with $\tan \beta$.
Therefore in the large $\tan \beta$ 
region one obtains the highest possible rates
for the direct DM searches. 
Similar results are presented in Ref.~\cite{Kim}.

For the correct calculation of the neutralino relic density  
in the large $\tan\beta$ region, 
 an unambiguous and
reliable determination of the $A$-mass is required.
The  details of the 
procedure in calculating the spectrum of the CMSSM can be
found elsewhere \cite{LS,LNSd2}. Here 
we shall only briefly 
refer to some subtleties which turn out to be
 essential for a correct determination of $m_A$.
In the CMSSM, 
$m_A$ is not a free parameter but 
is determined once the other parameters  are given.
$m_A$  depends sensitively on the Higgs
mixing parameter, $m_3^2$, which is determined from minimizing 
the  one-loop corrected effective potential.
For large $\tan \beta$ the derivatives of the effective potential
with respect the Higgs fields, which enter into the minimization conditions, 
are plagued by terms which are large and hence potentially dangerous, making
the perturbative treatment untrustworthy.
In order to minimize the large $\tan \beta$
corrections we had better calculate the effective potential using as
reference scale the average stop scale
$Q_{\tilde t}\simeq\sqrt{m_{{\tilde t}_1} m_{{\tilde t}_2} }$ \cite{scale}. 
At this scale these terms are small and hence perturbatively valid.
Also for the calculation of the pseudo-scalar Higgs boson mass
 all the one-loop corrections must be taken into account. 
In particular, the inclusion of  those of the neutralinos and charginos  
yields  a result for $m_A$ that is scale independent and 
approximates the pole mass to better than $2 \%$  \cite{KLNS}.
A more significant correction, which 
drastically affects the pseudo-scalar mass 
arises from the gluino--sbottom and chargino--stop corrections to the bottom
quark Yukawa coupling  $h_b$ \cite{mbcor,wagner,BMPZ,arno}.
The proper resummation of these corrections
is important for a correct determination of $h_b$ \cite{eberl,car2},
and accordingly of the $m_A$.
In calculating the $\lsp$
relic abundance, we solve the Boltzmann equation  
numerically using the machinery
outlined in Ref.~\cite{LNS}. In this calculation the coannihilation effects, 
in regions where $\tilde{\tau}_R$ approaches in mass the LSP, which is a high
purity Bino, are properly taken into account.
Seeking a precise determination  of the Higgs boson mass
the dominant two-loop corrections to this have been included \cite{zwirner}.
Concerning the calculation of the $\bsga$ branching ratio,
the important contributions beyond the leading order, 
especially for large $\tan \beta$ have been taken into account \cite{gamb}. 

\begin{figure}[t!] 
\begin{center}
\includegraphics[scale=1.25]{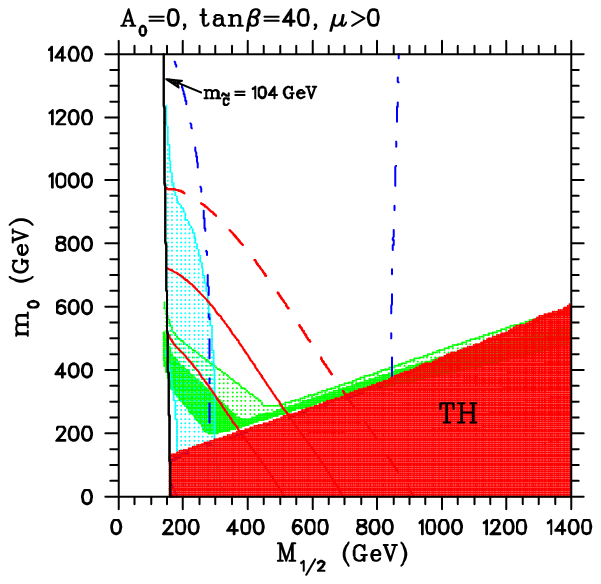}
\hspace*{.1cm}
\includegraphics[scale=1.25]{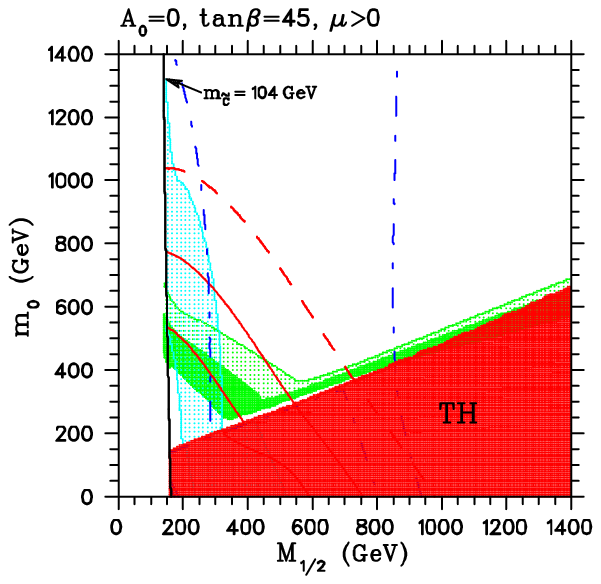}
\end{center}

\caption[]{
Cosmologically allowed regions of the relic density for 
 of $\tan \beta=40$ and $45$ in the $(M_{1/2},m_0)$ plane.  
The mass of the top is taken $175\GeV$. In the dark
green shaded area $0.08<\relic<0.18$. In the light green shaded
area $0.18<\relic<0.30\;$. The solid red lines mark the region within which
the supersymmetric contribution to the anomalous magnetic moment of the
muon is
$\alpha^{SUSY}_{\mu} = (361 \pm 106) \times 10^{-11}$.
The dashed red line
is the boundary of the region for which the lower bound is moved to
$2 \sigma$ limit. 
The dashed-dotted blue lines are
the boundaries of the region $113.5 \GeV \leq m_{Higgs} \leq 117.0 \GeV$.
The cyan shaded region on the right is excluded due to
$\bsga$ constraint. 
}

\label{fig1}  
\end{figure}

Using the new bound about the  $g_\mu-2$ as described in the
introduction, the parameter space constrained significantly.
This is shown in the next figures.  
In the panels shown in figure~\ref{fig1} we display our results by drawing 
the cosmologically
allowed region $0.08<\relic<0.18$ (dark green) in the $m_0, M_{1/2}$ plane 
for values of $\tan \beta$ equal to  $40$ and $45$ respectively.
Also drawn (light green) is the region $0.18<\relic<0.30$.
In the figures shown  we used for the top, tau and bottom masses
the values 
$M_t = 175 \GeV, M_{\tau} = 1.777\GeV$ and $m_b(m_b) = 4.25\GeV$. 
We have fixed $A_0=0$, since our results are not sensitive to
the value of the common trilinear coupling.
The solid red mark the region within which
the supersymmetric contribution to the anomalous magnetic moment of the
muon falls within the E821 range 
$ \alpha^{SUSY}_{\mu} = ( 361 \pm 106 ) \times 10^{-11}$.
The dashed red line
marks the boundary of the region when the more relaxed lower
bound $2\, \sigma$ value of the E821 range.
Along the blue dashed-dotted lines the light $CP$-even Higgs mass takes
values $113.5\GeV$ (left) and $117.0\GeV$ (right) respectively.
The line on the left
marks therefore the recent LEP bound on the Higgs mass \cite{LEP}.
Also shown is the chargino mass bound $104\GeV$.
The shaded area (in red)
at the bottom of each figure, labelled by TH, is theoretically disallowed 
since the light stau is lighter than the lightest of the neutralinos.
The cyan shaded region on the right is excluded by $\bsga$ constraint.

\begin{figure}[t] 
\begin{center}
\includegraphics[scale=1.25]{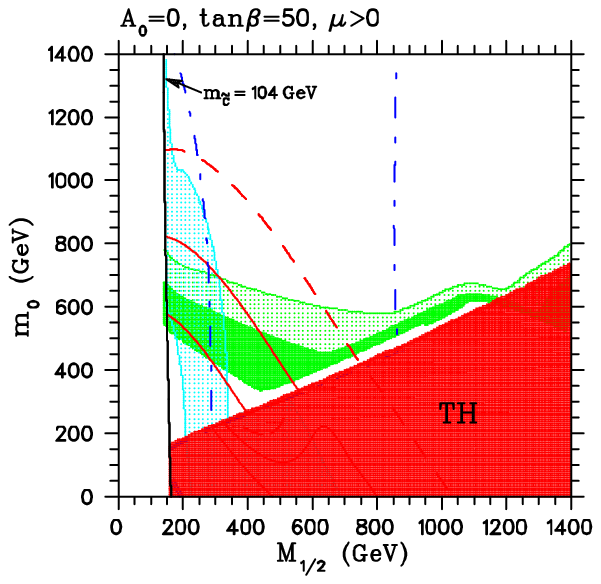}
\hspace*{.1cm}
\includegraphics[scale=1.25]{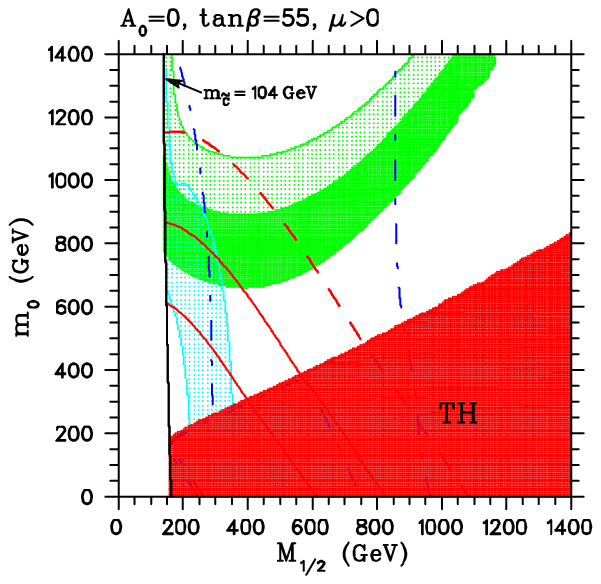}
\end{center}

\caption[]{The same as in Fig.~\ref{fig1} for  $\tan \beta=50$ and $55$. 
}

\label{fig2}  
\end{figure}

For large values of $\tan\beta$,
see the right panel of figure~\ref{fig2},
  a large region opens up within which the relic density
is cosmologically allowed. This is due to the pair annihilation of the
neutralinos through the pseudo-scalar Higgs exchange in the $s$-channel.
As explained before, for such high $\tan \beta$ the ratio $m_A / 2 m_{\lsp}$
approaches unity and the pseudo-scalar exchange dominates yielding
large cross sections and hence small neutralino relic densities. In this
case the lower bound put by the $g_\mu -2$ data 
cuts the cosmologically allowed
region which would otherwise allow for very large values of $m_0, M_{1/2}$.

\begin{table}[t]
\begin{center}
\begin{tabular}{|c|c|c|c|c|c|} \hline \hline
 $\tan\beta$ & $\lsp^0$ & $\tilde{\chi}^+$ & $\tilde{\tau}$ & $\tilde{t}$ & $h$ 
                                                                    \\ \hline
  15  &  128 (168) & 230 (304) & 150 (184) & 473 (637)  &  115 (116) \\
  20  &  155 (216) & 280 (390) & 174 (240) & 553 (838)  &  116 (118) \\
  30  &  164 (260) & 300 (470) & 186 (280) & 623 (986)  &  117 (118) \\
  40  &  161 (288) & 292 (520) & 256 (310) & 652 (1075) &  117 (119) \\
  50  &  218 (314) & 395 (568) & 430 (420) & 851 (1150) &  117 (119) \\
  55  &  154 (257) & 278 (466) & 450 (650) & 680 (988) &  115 (117) \\ 
 \hline \hline
\end{tabular}
\end{center}

\vspace{.4cm}
\caption{Upper bounds, in GeV, 
on the masses of the lightest of the neutralinos,
charginos, staus, stops and  Higgs bosons for various values of
$\tan\beta$ if the the E821 bounds are imposed.
The values within brackets represent the same situation when the $2 \sigma$
bound
$149 \times 10^{-11}<\alpha_{\mu}^{SUSY}<573 \times 10^{-11}$
is used.}

\label{table1}
\end{table}

For the $\tan \beta = 55$ case, 
close the highest possible value, and considering the 
$2 \, \sigma$ bound on the muon's anomalous magnetic moment
$\alpha_{\mu}^{SUSY} \geq 149 \times 10^{-11}$ and values of
$\relic$ in the range $0.13\pm0.05$, 
we find that the point with the highest value of $m_0$ is (in GeV) at
$(m_0, M_{1/2}) = (950, 300)$ and that with the highest value of 
$M_{1/2}$ is at $(m_0, M_{1/2}) = (600,750)$. 
The latter marks the lower end of
the line segment of the boundary $\alpha_{\mu}^{SUSY} = 149 \times 10^{-11}$ 
which amputates the cosmologically allowed stripe.
For the case displayed in the  right  panel of
the figure~\ref{fig2} 
the upper mass limits put on the LSP, and the lightest
of the charginos, stops and the staus are
$m_{\lsp} < 287, m_{{\tilde \chi}^{+}} <  539,  m_{\tilde t} < 1161,
m_{\tilde \tau} < 621$ (in $\GeV$). 
Allowing for $A_0 \neq 0$ values, the upper bounds put on $m_0, M_{1/2}$
increase a little and so do the aforementioned bounds on the sparticle
masses.
Thus it appears that the prospects of discovering CMSSM at  a
$e^{+} e^{-}$ collider with center of mass energy $\sqrt s = 800 \GeV$,
 are  {\em not} guaranteed. 
However in the allowed regions 
the next to the lightest neutralino, ${\tilde{\chi}^{\prime}}$, has a mass
very close to the lightest of the charginos and hence the process
$e^{+} e^{-} \goes {\tilde{\chi}} {\tilde{\chi}^{\prime}}$,
with 
${\tilde{\chi}^{\prime}}$ subsequently decaying to
$ {\tilde{\chi}} + {l^{+}} {l^{-}}$ or 
$ {\tilde{\chi}}+\mathrm{2\,jets}$,  
is kinematically allowed for such large $\tan \beta$, provided
the energy is increased to at least $\sqrt{s} = 900 \GeV$. It should
be noted however that this channel proceeds via the $t$-channel exchange
of a selectron
and it is suppressed
due to the heaviness of the exchanged
sfermion.

In table~{\ref{table1}} we give the upper bounds, in GeV,  
on the masses of the lightest of the neutralinos,
charginos, staus, stops and  Higgs bosons for various values of
$\tan\beta$ if the the E821 bounds are imposed.
The values within brackets represent the same situation when the weaker
$2 \sigma$ bound
$149 \times 10^{-11}<\alpha_{\mu}^{SUSY}<573 \times 10^{-11}$
is used. We take also into account the Higgs boson mass bound
as well as the $\bsga$ constrain.

   
\begin{figure}[t] 
\begin{center}
\includegraphics[scale=.7]{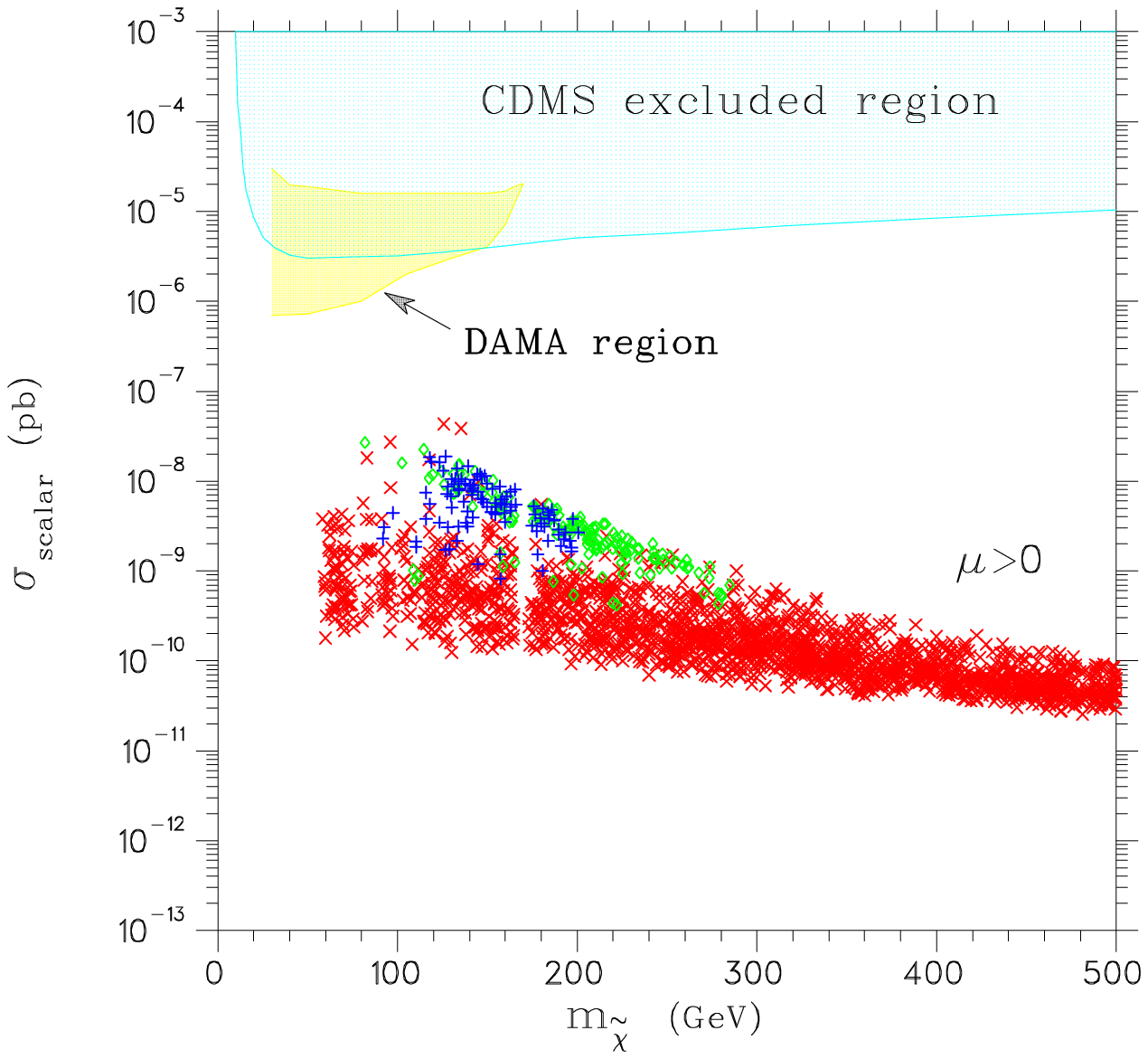}
\end{center}
\caption[]{
Scatter plot of
the scalar neutralino-nucleon cross section
 versus $\mlsp$, from a
random sample of 40,000 points.
On the top of the figure the CDMS excluded region and
the DAMA sensitivity region are illustrated. 
Blue pluses ($+$) are points within the E821 experimental region
$\almuon = ( 361 \pm 106 ) \times 10^{-11}$ and
which are cosmologically acceptable $\relic=0.13 \pm 0.05 $.
Green diamonds ($\diamond$)  are points within the $2\sigma$ 
E821 experimental region and cosmologically acceptable.
Crosses ($\times$) represent the rest of the random sample.
The Higgs boson mass bound $m_h > 113.5 \GeV$ and the $\bsga$
constraint are properly taken 
into account.}

\label{fig3} 
\end{figure}  

\section{Direct Dark Matter searches}
We turn now to study  the impact of the $g_\mu -2$, the $\bsga$ 
and the Higgs mass bounds
on the direct DM searches. For this reason we are using 
a random sample of 40,000 points in the region 
$|A_0|<1 \TeV$, $\tan \beta < 55$, $M_{1/2}<1.5 \TeV$, 
$m_0<1.5 \TeV$ and $\mu>0$.

In figure~\ref{fig3} we plot the scalar $\lsp$-nucleon   
 cross section as function of the LSP mass, $\mlsp$.
On the top of the figure the shaded region (in cyan colour) is
excluded by the CDMS experiment \cite{cdms}.
The DAMA sensitivity region (coloured in yellow) is
also plotted \cite{dama}. 
Pluses ($+$) (in blue colour) represent points 
which are both compatible
 with the E821 data $\almuon = (361 \pm 106)\times 10^{-11}$
and the cosmological bounds for the neutralino relic density
$\relic = 0.13 \pm 0.05$. 
Diamonds ($\diamond$) (in green colour) represent points 
which are  compatible
 with the $2 \, \sigma$ E821 bound 
$149 \times 10^{-11}  < \almuon  < 573 \times 10^{-11} $
and the cosmological bounds.
The crosses ($\times$) (in red colour) represent the rest of the
points of our random sample.  Here the Higgs boson
mass, $m_h > 113.5 \GeV$ and the
$\bsga$ bounds have been properly taken into account.
From this figure it is seen that the     
points which are compatible both the ($g_{\mu}-2$) E821
and the cosmological data (crosses) yield cross sections
of the order of $10^{-9}$ pb and  the maximum value of  the $\mlsp$
is about $200 \GeV$.
Accepting the $2\, \sigma$ $g_{\mu}-2$ bound we get 
cross sections
of the order of $10^{-10}$ pb and $\mlsp \sim 300\GeV$.
\begin{figure}[t] 
\begin{center}
\includegraphics[scale=.7]{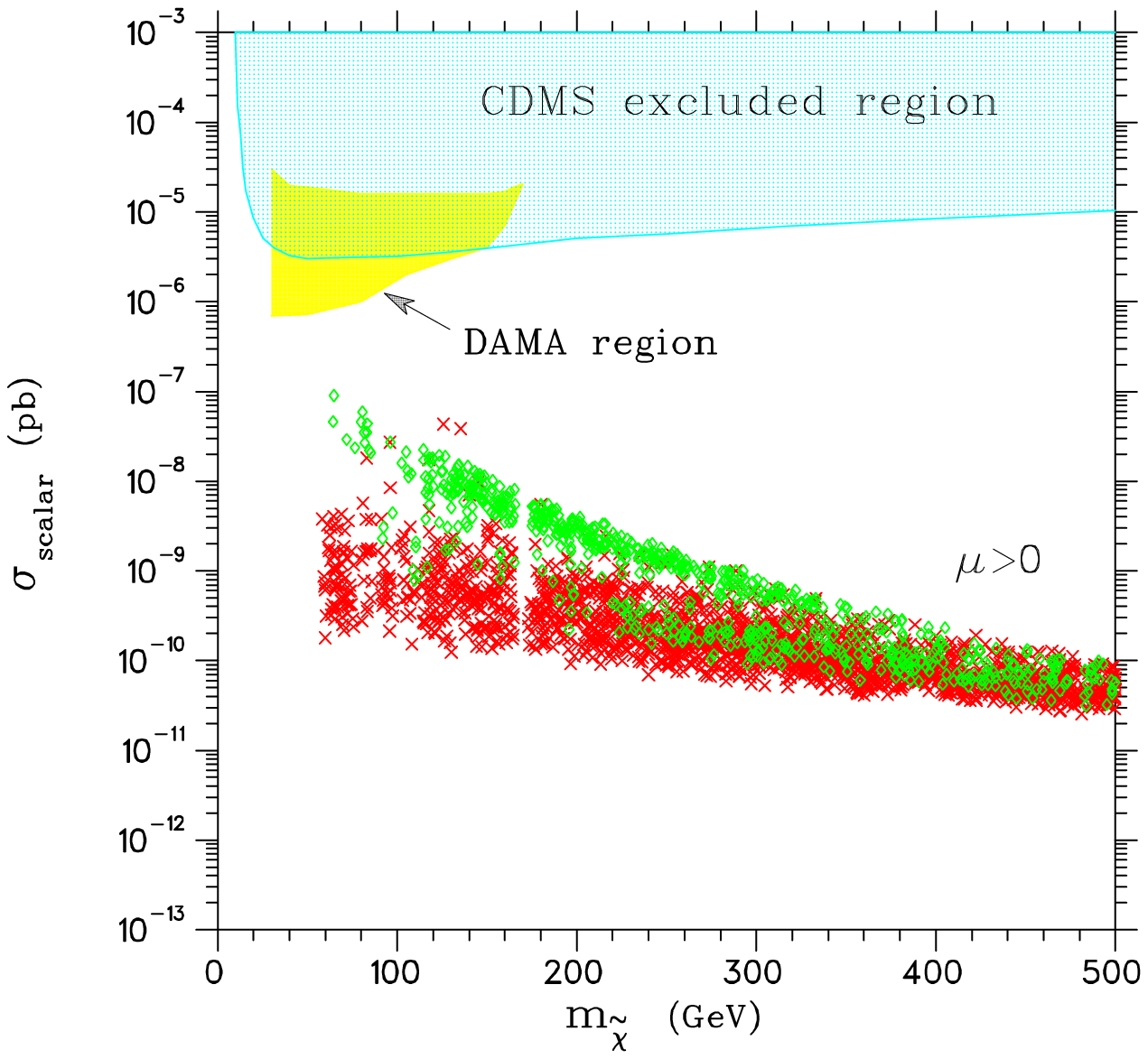}
\end{center}
\caption[]{
Scatter plot of
the scalar neutralino-nucleon cross section
 versus $\mlsp$, from a
random sample of figure~\ref{fig2}.  
Diamonds ($\diamond$) are  cosmologically
acceptable points, without putting any restriction from the
$\almuon$. Crosses ($\times$) represent points with
unacceptable $\relic$.
The Higgs boson mass bound $m_h > 113.5 \GeV$ and the $\bsga$
constraint are taken into account.}

\label{fig4} 

\end{figure}  

In figure~\ref{fig4} we don't impose the constraints stemming
from $g_{\mu}-2$ data, therefore  due to the coannihilation processes
the cosmologically acceptable LSP mass can be heavier than
$500 \GeV$.  It is worth    noticing that
 that imposing the $g_{\mu}-2$ bound  
the lowest allowed $\lsp$-nucleon cross section increases by about
one order of magnitude, from $10^{-11}$ pb to $10^{-10}$ pb.
Considering the $\mu>0$ case, it is important that
using the  cosmological bound for the DM alone,  one can put 
a lower bound on $\sxsec \simeq 10^{-11}$.
This fact is very encouraging for the future
DM direct detection experiments~\cite{Klapdor}.
Unfortunately this is not the case for $\mu<0$, where
the  $\sxsec$ can be very small, due to an accidental
cancellation between the sfermion and Higgs boson exchange processes.
However, this case is not favoured by $g_\mu-2$ and also  $\bsga$ data.

\section{Conclusions}
In conclusion, 
we have combined recent high energy physics experimental information,
like the anomalous magnetic moment of the muon measured at
 E821 Brookhaven experiment,
the $\bsga$ branching ratio  and the light Higgs boson mass bound
from LEP, with the  cosmological data for DM.
Especially for the $g_\mu-2$ bound 
we have used the very recent value from the E821 experiment
along with a new determination of the 
hadronic vacuum polarization contribution
to the anomalous  magnetic moment of the muon. 
We studied the imposed constraints on the
parameter space of the CMSSM and hence we assessed the potential
of discovering SUSY, if it is based on CMSSM, at future colliders
and DM direct searches experiments.
The use of the $2\, \sigma$ $g_\mu-2$ bound
 can guarantee that in LHC  but also in 
a $e^{+}e^{-}$ linear collider with center of mass energy
$\sqrt{s} = 1200\GeV$,  CMSSM can be discovered.
 The effect of these constraints is  also significant 
for the direct DM searches.
For the $\mu>0$ case
we found that the
minimum
value of the spin-independent $\lsp$-nucleon
cross section attained
is of the order of $10^{-10}$ pb.

\vspace*{1cm}
\noindent 
{\bf Acknowledgements} \\ 
\noindent A.B.L. acknowledges support from HPRN-CT-2000-00148
and HPRN-CT-2000-00149 programmes. He also thanks the University of Athens
Research Committee for partially supporting this work. 
D.V.N. acknowledges support by D.O.E. grant DE-FG03-95-ER-40917.
V.C.S. acknowledges support  by a Marie Curie Fellowship of the EU
programme IHP under contract HPMFCT-2000-00675.
The authors  thank  the Academy of Athens for supporting this work.

\end{document}